\documentclass{emulateapj}\usepackage{apjfonts}
\usepackage{graphicx}
\def\chan{{\sl Chandra}}
\def\psr{J1357--6429}

\def\xmm{{\sl XMM}-Newton}
\def\pd{\dot{P}}
\def\ed{\dot{E}}
\def\ergs{ergs s$^{-1}$}
\def\ergss{ergs cm$^{-2}$ s$^{-1}$}
\def\fp{p_f}
\def\ls{\lower 2pt \hbox{$\;\scriptscriptstyle \buildrel<\over\sim\;$}}
\def\gs{\lower 2pt \hbox{$\;\scriptscriptstyle \buildrel>\over\sim\;$}}
\def\ref{\par \hangindent 10 pt\noindent}
\def\tc{\tau_{\rm c}}
\def\nh{n_{\rm H,21}}
\def\dd{d_{2.5}}

\shorttitle{PSR~J1357--6429 in X-rays}
\shortauthors{V. E. Zavlin}

\begin{document}

\title{First X-ray Observations Of The Young Pulsar J1357--6429}

\author{Vyacheslav E. Zavlin}
\affil{Space Science Laboratory, NASA MSFC SD59, Huntsville, AL 35805;
vyacheslav.zavlin@msfc.nasa.gov}

\begin{abstract}
The first short \chan\ and \xmm\ observations of the young and
energetic pulsar \psr\ provided 
indications of a tail-like
pulsar-wind nebula associated with this object, as well as 
pulsations
of its X-ray flux with a pulsed fraction $\fp\ga50$\%
and a thermal component dominating at 
energies $E\la2$ keV.
The elongated nebula is 
very compact in size
and might be interpreted as evidence
for a pulsar jet.
The thermal radiation is most plausibly emitted from the entire neutron star
surface of 
a 10 km radius and
a $1.0\pm 0.1$ MK temperature, covered with a
hydrogen atmosphere. At higher energies the pulsar's emission
is of a nonthermal 
origin, with a power-law spectrum
of a photon index $\Gamma=1.1\pm 0.2$. This makes the 
properties of 
PSR~\psr\ very similar to those of the 
young pulsars J1119--6127
and Vela with detected thermal radiation. 
\end{abstract}


\section{Introduction}
Once discovered in radio, young rotation-powered pulsars
with characteristic ages of $\tc=P/2\pd\sim 10$ kyr
($P$ is pulsar spin period)
often become targets for X-ray observations
because these objects usually possess
large values of the rotational-energy loss, 
$\ed\gs 10^{36}$ \ergs,
and hence are expected to generate observable nonthermal (magnetospheric) 
emission and
power detectable pulsar-wind nebulae 
(PWNe; see Kargaltsev, Pavlov, \&
Garmire~2007 for 
properties of a number of PWNe). 
Measuring 
parameters of  
the magnetospheric radiation is particularly
important for inferring emission processes operating in pulsar 
magnetosphere.
Studying 
a PWN is
essential for determining the energy of the pulsar wind, probing
the ambient medium, and understanding the
shock acceleration mechanism(s).

On the other hand, neutron stars (NSs), born hot in supernova
explosions, cool down via neutrino emission from the entire
NS body and
heat transport through the stellar envelope to the surface
and subsequent thermal 
radiation.
The NS cooling rate 
is determined by various 
processes
of the neutrino emission, as well as the (unknown) 
properties of the NS inner matter
(Yakovlev \& Pethick 2004).
Till recently, 
young pulsars were thought
to be
such powerful nonthermal X-ray emitters that
their thermal radiation 
would be completely buried under
the nonthermal component.
However, 
new observations 
revealed 
the pulsars Vela ($\tc=11.3$ kyr) and J1119--6127 ($\tc$$=$1.6\,kyr)
as examples of young NSs with 
the bulk of the X-ray flux 
of a thermal origin (Pavlov et al.~2001; Gonzalez et al.~2005).
The surface emission is of a special interest because
confronting observational
data with model 
thermal radiation, presumably formed
in NS atmospheres, can allow one to infer the surface temperatures and
magnetic fields, 
constrain the NS mass ($M$) and radius ($R$), and
understand the 
properties of the superdense matter in the
NS interior (see Zavlin~2007a [hereafter Z07] for 
a review).

At a distance of $d\simeq 2.5$ kpc
derived from the pulsar dispersion measure 
and the 
Galactic electron density model
of Cordes \& Lazio~(2002),
PSR~\psr\ ($P=166$ ms, $\tc=7.3$ kyr, 
$\ed=3.1\times 10^{36}$ \ergs, 
surface magnetic field $B\sim8\times10^{12}$ G), 
discovered by Camilo et al. (2004; hereafter C04)
during the Parkes multibeam survey of the Galactic plane,
is one of the nearest 
objects in the group of about two dozen currently known
pulsars\footnote{{\tt http://www.atnf.csiro.au/research/pulsar}}
with $\tc<15$ kyr.
The actual pulsar's age could be 
$\tau=2\tc/[n-1]\ls 15$ kyr, 
for a typical value of the magneto-dipole braking index $n\simeq 2$--3.
Radio data on this pulsar 
(C04) revealed a strong
glitch of its spin period, $\Delta{P}/P\simeq -2.4\times 10^{-6}$,
similar to those experienced by some other young pulsars (e.~g., Vela),
supporting the hypothesis that PSR~\psr\ is 
a young NS.
It is located relatively close to the 
supernova remnant (SNR) candidate G309.8--2.6,
but a possible connection between these two objects
has yet to be established
(see C04 for details),
opposite to the case of the majority of young pulsars associated with 
SNRs.

As with many other young pulsars, PSR~\psr\ has been been a target of
both \chan\ and \xmm. Sections\,2 and 3 
describe these observations 
and analysis of the collected X-ray data. 
The results are discussed in Sec.\,4. 

\section{Observations and data reduction}
\xmm\ observed PSR~\psr\ on 2005 August 5 
for 11.6 and 14.5 ks
of effective exposures with the European Photon Imaging Cameras, 
EPIC-pn and MOS (respectively),
operated in Full-Frame Window mode.
The highest achieved time resolution of 73.4 ms 
is not sufficient for timing of a signal of
a 166-ms period. The data were reprocessed with the \xmm\ Science Analysis
Software (\texttt{SAS} v.~7.0.0).
Due to strong particle flares a background
level during 
about 85\% of the total observational span was
higher by a factor of 3--20 
than the ``quiescent'' rate.

The pulsar was observed with the
\chan\ High Resolution
Camera (HRC-S)
operated in Timing mode
on 2005 November 18 and 19,
for 16.1 and 17.1 ks of effective exposure 
and with a 16-$\mu$s time resolution. 
The \chan\ Interactive Analysis of Observations
(\texttt{CIAO}) software (v.~3.4; \texttt{CALDB} ver.3.3.0.1)
was used to generate the level~2 event files 
from the two observations. Those 
were combined with the 
\texttt{CIAO} script \texttt{`merge\_all'}, that resulted in the
data product of a 33.2 ks exposure.
The pulse-invariant (PI) channels 
with PI$<$25 or PI$>$150 were
filtered out 
to minimize background contamination.  

\section{Observational results}
\subsection{Spatial analysis}
Figure\,1 presents a smoothed HRC-S image of the $16''\times 16''$ region 
around the radio position of PSR~\psr\ derived by C04.
The image reveals a point-like
source surrounded by a weaker diffuse emission.
The source is centered at the position
R.A.$=13^{\rm h} 57^{\rm m} 2\fs54$ and Dec.$=-64^\circ 29'' 30\farcs0$.
The difference of $0\farcs74$ between the 
X-ray and radio positions is 
very close to the 1$\sigma$ error in the 
\chan\ HRC-S positional astrometry\footnote{
{\tt http://asc.harvard.edu/cal/ASPECT/celmon}}. The diffuse emission
feature, extending toward northeast to a distance of about $2''$ from the
center of the point-like source, is seen in the both HRC-S datasets and
is not an image artifact caused by the instrument anomalies
and/or background\footnote{
{\tt http://cxc.harvard.edu/proposer/POG/html/HRC.html}}.
The presence of the extended emission is also apparent in Fig.\,2 
showing two 1-D distributions of counts
in rectangular apertures 
shifted along the main symmetry axis of the box plotted in Fig.\,1:
one histogram presents the distribution of detected counts,
the other indicates the HRC-S point-spread function computed 
from a 2-D simulation performed with
the \texttt{CIAO} \texttt{'mkpsf'} tool.
The proximity of the X-ray and radio positions and the morphology 
of the extended feature strongly suggest that the detected X-ray radiation
is emitted by PSR~\psr\ and its possible PWN\footnote{
No other point-like source as a possible counterpart of this feature
is found in available observational catalogues.}.

\begin{figure}
\includegraphics[width=\hsize]{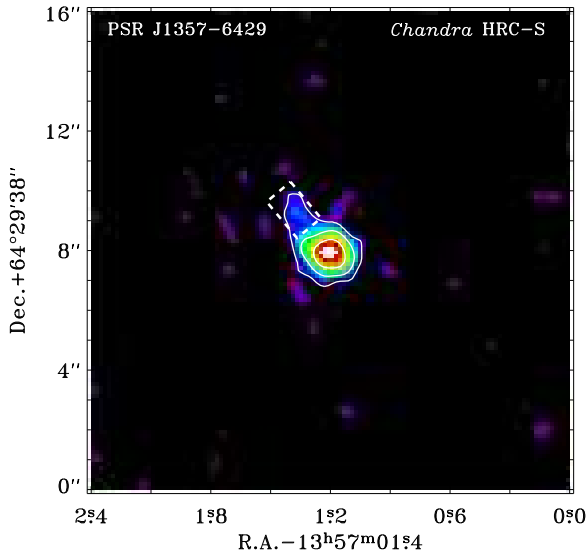}
\caption{\chan\
HRC-S $16''\times 16''$ image smoothed with a $0\farcs4$ FWHM Gaussian
showing PSR~\psr\ surrounded by an extended structure
elongated in the east-north direction (indicated by the white-dashed
box of a $1''\times 1\farcs5$ size).
White 
contours correspond to intensity values of
0.17, 0.55, and 1.74 counts arcsec$^{-2}$ ks$^{-1}$.
}
\label{fig1}
\end{figure}

\begin{figure}
\includegraphics[height=5cm]{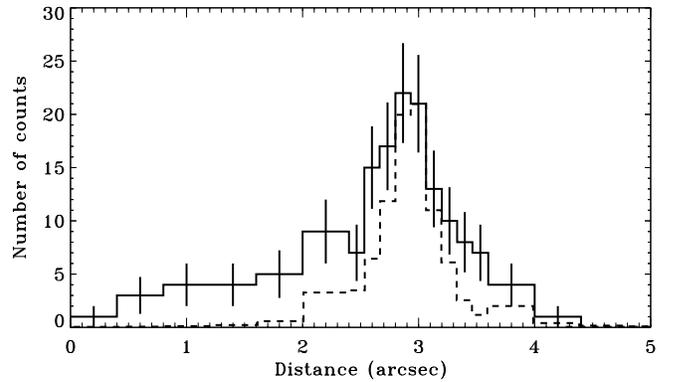}
\caption{
Distribution of counts detected with the \chan\ HRC-S 
instrument (solid histogram)
in rectangular apertures with a $1\farcs84$ height 
and varying widths ($0\farcs13$ or $0\farcs40$) shifted along the main
symmetry axis of the box shown in Fig.~1. The dashed histogram shows the
HRC-S point-spread function computed for the same apertures.
}
\label{fig2}
\end{figure}

The $1\farcs05\times 1\farcs58$ box shown in Fig.\,1 contains 18
counts, of them only 14\% belongs to background. Hence,
the elongated ``tail'' of the diffuse emission (within the 
box) emits at a rate of $0.48\pm 0.14$ counts ks$^{-1}$.
Modeling the tail's emission with a power-law (PL) spectrum of a photon index
$\Gamma=1.5$ (typical for X-ray PWNe --- e.~g., Kargaltsev et al.~2007) 
absorbed with the hydrogen column density
$\nh=n_{\rm H}/(10^{21}\,{\rm cm}^{-2})=5.0$ (see Sec.\,3.3)
yields the tails's 
(unabsorbed) flux 
$F_{\rm t}\simeq 3.3\times 10^{-14}$ \ergss
in 0.5--10 keV,
or the luminosity
$L_{\rm t}\simeq 2.5\times 10^{31}\dd^2$ \ergs
($\dd=d/[2.5\,{\rm kpc}]$).

No large-scale signatures of a diffuse emission
which could be associated with the SNR candidate G309.8--2.6 
(Sec.\,1) were found in 
these data. 
An upper limit on the averaged surface brightness of the possible
SNR emission, estimated from the $2\farcm5$-radius circle centered
at the position 
R.A.$=13^{\rm h} 56^{\rm m} 57\fs0$ and Dec.$=-64^\circ 28'' 00$,
is 5 counts ks$^{-1}$ arcmin$^{-2}$.

\subsection{Timing analysis}
For the timing analysis, 137 counts were extracted from the
$1\farcs05$-radius circle centered at the pulsars X-ray position, 
with 132 counts 
estimated to belong to
the pulsar. The photon times of arrival 
were corrected to the solar system barycenter using the
\texttt{CIAO} \texttt{'axBary'} tool
and the pulsars radio position and 
convolved with the accurate timing solution
obtained from radio observations of the pulsar conducted
at the Parkes telescope\footnote{
The Parkes Observatory is part of the Australia Telescope, which is funded
by the Commonwealth of Australia for operation as a National Facility
managed by CSIRO.}
in the interval
from 2005 April 27 to 2005 December 10 (53,487--53,714~MJD):
$f= 6.019298216(5)$ Hz and, $\dot{f}=-1.29968(6)\times 10^{-11}$ Hz s$^{-1}$ 
at the epoch 53,694.0 MJD (given with the 1$\sigma$ uncertainties;
F. Camilo 2007, private communication). This radio solution leaves
a timing residual of only 3.6 ms (2.2\% of the spin period).
At this ephemeris, the $Z^2_m$--test ($m$ is number of harmonics) 
results in the value of $Z^2_1=12.0$ ($Z^2_2=14.6$), or 
a signal detection at a 99.752\%
confidence level. 

Figure\,3 presents
the pulse profile of PSR~\psr\ extracted at the radio spin parameters,
with the estimated pulsed fraction, $\fp=63\pm 15$\%
(defined as the fraction of counts above the minimum
in the light curve).
The $\chi^2$--test yields the value $\chi^2_\nu=4.07$ for $\nu=9$ d.o.f,
meaning a 95.996\% probability that the pulse profile 
differs from a constant line. 
Because of the poor statistics, the shape of the pulse profile
is not well determined. The only rather
certain fact is that the pulsed fraction is large,
implying special properties of the detected X-ray emission (see Sec.\,4).

\begin{figure}
\includegraphics[height=6cm]{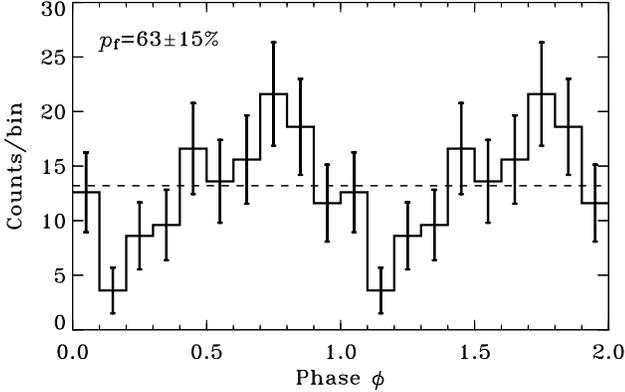}
\caption{
Pulse profile of PSR~\psr\ extracted from the \chan\
HRC-S data. 
The dashed line indicates the mean level.
}
\label{fig3}
\end{figure}

\begin{figure}
\includegraphics[height=9cm]{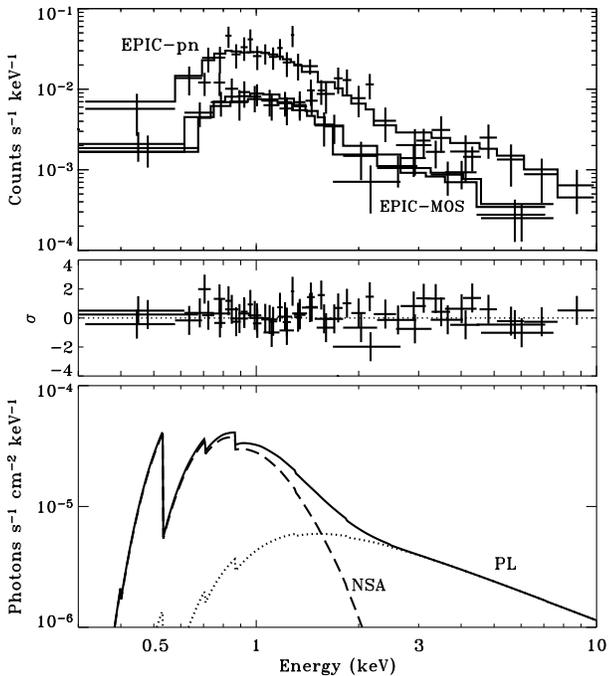}
\caption{
Spectra of PSR~\psr\ detected with 
the 
\xmm\ EPIC instruments
and fitted with a two-component, NS atmosphere (NSA) plus power law
(PL), model.
}
\label{fig4}
\end{figure}

\subsection{Spectral analysis}
To optimize the signal-to-noise ratio,
the EPIC
spectra were extracted 
from $20''$-radius circles centered at the pulsar position
(these 
include
the emission of the faint PWN 
suggested by the HRC-S data).
The extraction regions contain about 75\% of the source flux.
The instrumental responses were generated with the \texttt{SAS}
\texttt{`rmfgen'} and \texttt{`arfgen'} tools.
Background was evaluated from similar regions in vicinity of the
pulsar position. The estimated source count rates (corrected for the
finite extraction aperture) are $56\pm 5$ and $17\pm 2$ counts ks$^{-1}$
in 
0.3--10 keV,
for EPIC-pn and MOS, respectively.

A single PL model fits the
spectra rather well,
with ${\rm min}\,\chi^2_\nu=1.17$ for $\nu=59$ d.o.f, yielding
$\Gamma=2.1\pm 0.4$ and $\nh=2.3\pm 0.8$
(1$\sigma$ errors are given here and below).
The derived nonthermal (isotropic) luminosity in 
0.5--10 keV 
is
$L_{\rm nonth}\simeq 2.0\times 10^{32}\dd^2$\,\ergs$=6.5\times 10^{-5}\ed\dd^2$
(of this, about 10\% belongs to the PWN). 
On the other hand,
it is 
reasonable to assume that PSR~\psr\ emits observable thermal radiation.
Adding a thermal component 
improves the fit, ${\rm min}\,\chi^2_\nu$$=$0.76 for $\nu$$=$57 d.o.f, 
indicating that
it is required by the data at a 99.9998\% (4.8$\sigma$) confidence 
level.
The thermal component can be equally well described with both a
blackbody (BB) spectrum and a NS magnetized
($B=1\times 10^{13}$\,G) hydrogen atmosphere model\footnote{
The
\texttt{`nsa'} code in the XSPEC 
package.}.
However, as in many other cases, the estimates
on the temperature and size of the emitting area 
derived from
these two models are very different(see Zavlin~2007a for details):
$T^\infty_{\rm bb}=1.7\pm 0.2$ MK and
$R^\infty_{\rm bb}=(2.5\pm 0.5)\dd$ km (redshifted values) in the BB fit,
and $T_{\rm eff}=1.0\pm 0.1$ MK (effective temperature)
for $R=10$ km at $\dd=1$ in the NS
atmosphere fit.
The corresponding
bolometric luminosities are 
$L_{\rm bb}^\infty\simeq 3.7\times 10^{32}\dd^2$ \ergs\
and $L_{\rm atm}\simeq 7.1\times 10^{32}$ \ergs\
($L_{\rm atm}^\infty=[1-2GM/c^2R] L_{\rm atm}\simeq 0.6 L_{\rm atm}$
for $M=1.4 M_\odot$ and $R=10$ km).
The parameters of
the nonthermal component are similar, 
$\Gamma=1.3\pm 0.2$ and $1.1\pm 0.2$, in the BB and atmosphere fits
(respectively), with 
$L_{\rm nonth}\simeq 1.4\times 10^{32}\dd^2$\,\ergs$=4.5\times 10^{-5}\ed\dd^2$.
The inferred hydrogen column density,
$\nh=4.9\pm 2.0$, is somewhat larger than in the single-PL interpretation
but still consistent with the standard estimate
suggested by the pulsar dispersion measure ${\rm DM}=127.2$ cm$^{-3}$ pc,
$n_{\rm H,DM}/(10^{21}\,{\rm cm}^{-2})={\rm DM}/(10^{20}\,{\rm cm}^{-2})
\simeq 4.0$ (assuming a 10\% ionization degree of the interstellar hydrogen), 
and is
reasonably lower than the galactic neutral hydrogen density estimated in the
direction to PSR~\psr, $n_{\rm HI}/(10^{21}\,{\rm cm}^{-2})\simeq 10.0$.
It is also worth mentioning that the thermal componet contributes 72\% of the
X-ray flux observed at $E<2$ keV, regardless of whichever model is used to
describe the component.
The best two-component fit involving the NS atmosphere model is
shown in Fig.\,4.

\section{Discussion}
The first X-ray observations of PSR~\psr\ 
have revealed interesting properties of this object\footnote{
Esposito et al. (2007)
analyzed the same data,
obtained similar results in the PL and BB-plus-PL spectral fits
and put an upper limit $\fp<30\%$
for a possible sinusoidal
signal. They searched for a diffusion emission of a $10''$ size and
failed to notice the small extended feature.},
including the indication of the compact tail-like PWN
associated with the pulsar.
The estimated efficiency with which PSR~\psr\ powers 
the PWN,
$L_{\rm t}/\ed\approx 0.8\times 10^{-5}$, 
can be considered as ``medium'' among those measured for
tail-like PWNe associated with other pulsars. For example,
it is lower by about an order of magnitude than the efficiencies
of the PWNe of the young pulsar B1757--24 (Kaspi et al.~2001) and
the millisecond pulsar B1957+20 (Stappers et al.~2003; Zavlin~2007b)
but larger by a factor of a few than those inferred for the Geminga's
tail (Pavlov, Sanwal, \& Zavlin~2006) and the Vela's southest jet
(Pavlov et al.~2003). On the other hand, the derived luminosity of 
the 
PWN associated with PSR~\psr\ is about by a factor
of 5--8 smaller than the nonthermal luminosity emitted by the pulsar
itself, that does not match the correlation,
$L_{\rm PWN}\approx 5 L_{\rm nonth}$, inferred by Kargaltsev et al.~(2007)
for a sample of pulsars.  
It is premature to make 
firm conclusions about
the properties of the possible PWN.
However, 
despite the fact that
the proper motion of PSR~\psr\ has yet not been measured,
the shape and estimated luminosity
of this elongated feature may suggest a 
hypothesis that
it is a pulsar jet along
the pulsar spin axis aligned with its velocity,
similar to the interpretation proposed for
the tail-like features of the Crab,
Vela, and Geminga pulsars
(Weisskopf et al.~2000; Pavlov et al.~2003; Pavlov
et al.~2006).
On the other hand, if the elongation direction and that of the
proper motion turn out to be perpendicular to each other,
it would indicate that the X-ray emission of the PWN emerges
from a torus associated with the equatorial pulsar wind. 
Future measurement of the pulsar proper
motion is important for validating this hypotheses.

\begin{figure}
\includegraphics[height=6cm]{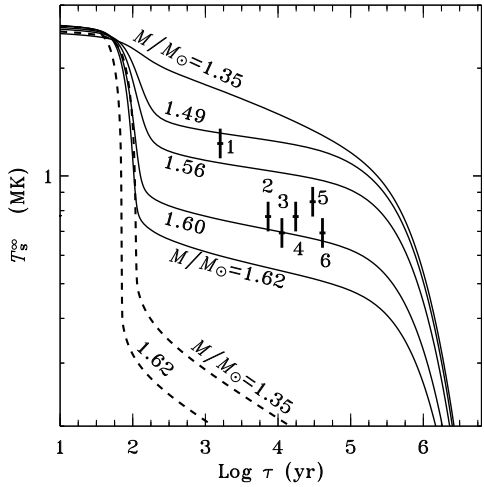}
\caption{NS cooling models (Yakovlev \& Pethick~2004)
with the 
`2p' proton superfluidity  (solid curves)
and without the superfluidity (dashed curves), for 
different NS masses.
The crosses indicate the pulsars:
J1119--6127 (1), 
J1357--6429 (2),
Vela (3), B1706-44 (4), J0538+2817 (5),
and B2334+61 (6). The redshifted surface temperature is 
$T_{\rm s}^\infty=[1-2GM/c^2R]^{1/2} T_{\rm eff}$ ($=0.77 T_{\rm eff}$
for $M/M_\odot=1.4$ and $R=10$ km), 
with
$T_{\rm eff}$ 
being the effective 
temperature of the NS surface
obtained
in the fits 
with the NS atmosphere models.
}
\label{fig5}
\end{figure}

The spectral data on PSR~\psr\ show 
that the bulk of the pulsar X-ray flux is probably of a thermal origin.
It makes this object the second youngest NS in the group of 
energetic 
pulsars 
with a thermal 
component dominating at 
$E\la 2$ keV.
In addition to PSRs~J1119--6127, \psr\ and Vela, the group also includes
PSRs B1706--44 with $\tc=17.5$ kyr, J0538+2817 with
the derived true age $\tau\simeq\tc/20.5=30$ kyr
(Kramer et al.~2003), and B2334+61 with $\tc=40.9$ kyr 
(see 
Z07 for details).
Similar to the situation
with 
other members of
this group, the origin of thermal component of PSR~\psr\ cannot be
unambiguously determined from the spectral data alone.
The parameters of the thermal
component obtained 
in the BB fit may be 
interpreted as radiation from a small hot area (polar caps) on the 
NS surface,
although the estimated radius, $R^\infty_{\rm bb}\simeq 2.5$ km,
is much greater than the canonical polar cap estimate,
$R_{\rm pc}=(2\pi R^3/cP)^{1/2}\simeq 0.3$ km.  
Opposite to this,
the fit with the magnetized NS atmosphere models indicate
that the thermal emission originates from the entire NS surface.
One could discriminate between these two interpretations invoking
pulsations of the X-ray emission. In the case of PSR~\psr,
as the \chan\ HRC-S instrument is most efficient at $E\la 2$ keV,
the pulsed fraction of
$\fp\ga 50$\%
suggests that the thermal emission is
intrinsically anisotropic, as predicted by the NS atmosphere models ---
otherwise the effect of strong gravitational bending of photon
trajectories 
near the NS surface would strongly suppress the pulsations.
This obviously contradicts the simplistic (isotropic) BB interpretation.
Such a pulsed fraction also indicates
that the NS has a strong nonuniformity of
the surface temperature and magnetic field. Hence, the temperature
$T_{\rm eff}$ inferred from the NS atmosphere fit assuming the
uniform surface should be considered as an approximate estimate on 
the ``mean'' surface temperature.  
The same conclusions stand for 
PSR~J1119--6127 with a pulsed fraction $\fp\ga 60$\%
(Gonzalez et al.~2005),
as well as PSR~0538+2817 whose pulse profile of the thermal flux 
emission shows a complicate dependence on photon energy (Zavlin \& Pavlov 2004).

Assuming that the characteristic ages of the five pulsars of this
group are close to
the true ones\footnote{This assumption should be taken with caution ---
see the case of PSR~J0538+2817.}, it is illustrative to compare
NS cooling models with
the surface temperatures estimated for these objects from 
the 
NS atmosphere 
fits.
Figure\,5 presents cooling models
with and without proton (`2p' model) superfluidity in the NS core 
(Yakovlev \& Pethick~2004), as well as the measured temperatures.
The superfluidity reduces the neutrino emission by suppressing
the Urca processes and, hence, decelerates the NS cooling.
This effect depends on NS mass, being stronger for higher masses.
The comparison
suggests that the interiors of these six pulsars are superfluid,
and the NS masses may be in the range $M=[1.5$--$1.6] M_\odot$
(note that this range would be different for another
model superfluidity). 

However, despite the important X-ray information on PSR~\psr\
available,
the pulsar's 
properties
remain yet rather 
uncertain
because of
the scanty number of photons detected 
in these short 
exposures.
Therefore, deeper observations are required
to provide data of much better quality to firmly establish
the actual shape, size, and spectrum of the possible PWN, and
accurately infer the X-ray spectral and temporal
properties of the pulsar.

\acknowledgements{}
The author thanks 
F. Camilo for providing the pulsar radio ephemeris 
and remarks to the manuscript,
D. Yakovlev for making available the NS cooling models,
and A. Tennant and M. Weisskopf for useful discussions.
This work is supported by a NASA Senior Associateship
Award at the NASA MSFC.

\end{document}